\documentclass[11pt,a4paper]{article}

\usepackage[T1]{fontenc}
\usepackage[utf8]{inputenc}
\usepackage{authblk}

\usepackage{lipsum}
\usepackage{amsfonts}
\usepackage{graphicx}
\usepackage{epstopdf}
\usepackage{algorithmic}
\usepackage{epsfig,color} 
\usepackage{eufrak}

\usepackage{array}
\usepackage{geometry,xcolor}        
\usepackage{amssymb,amsmath}
\usepackage{dsfont}
\usepackage{amsthm} 
\usepackage{hyperref} 

\usepackage[caption=false]{subfig} 

\newcommand{\dd}{\mathrm{d}}
\newcommand{\gfp}{$\mathcal G$}
\newcommand{\R}{\mathbb{R}}

\def\Fref#1{Figure~\ref{#1}}
\def\fref#1{Figure~\ref{#1}}

\def\tref#1{Table~\ref{#1}}

\begin{document}

\title{Noisy network attractor models for transitions between EEG microstates}

\author[1,*]{Jennifer Creaser}
\author[1]{Peter Ashwin}
\author[2]{Claire Postlethwaite}
\author[3]{Juliane Britz}
\affil[1]{Department of Mathematics and EPSRC Centre for Predictive Modelling in Healthcare, University of Exeter, Exeter, Devon, UK}
\affil[2]{Department of Mathematics, University of Auckland, Auckland, NZ}%
\affil[3]{Department of Psychology and Department of Medicine, Neurology, University of Fribourg, Fribourg, Switzerland}%
\affil[*]{j.creaser@exeter.ac.uk}

\date{}

\maketitle


\begin{abstract}
The brain is intrinsically organized into large-scale networks that constantly re-organize on multiple timescales, even when the brain is at rest. The timing of these dynamics is crucial for sensation, perception, cognition and ultimately consciousness, but the underlying dynamics governing the constant reorganization and switching between networks are not yet well understood. Functional magnetic resonance imaging (fMRI) and electroencephalography (EEG) provide anatomical and temporal information about the resting-state networks (RSNs), respectively. EEG microstates are brief periods of stable scalp topography, and four distinct configurations with characteristic switching patterns between them are reliably identified at rest. Microstates have been identified as the electrophysiological correlate of fMRI-defined RSNs, this link could be established because EEG microstate sequences are scale-free and have long-range temporal correlations. This property is crucial for any approach to model EEG microstates. This paper proposes a novel modeling approach for microstates: we consider nonlinear stochastic differential equations (SDEs) that exhibit a noisy network attractor between nodes that represent the microstates. Using a single layer network between four states, we can reproduce the transition probabilities between microstates but not the heavy tailed residence time distributions. Introducing a two layer network with a hidden layer gives the flexibility to capture these heavy tails and their long-range temporal correlations. We fit these models to capture the statistical properties of microstate sequences from EEG data recorded inside and outside the MRI scanner and show that the processing required to separate the EEG signal from the fMRI machine noise results in a loss of information which is reflected in differences in the long tail of the dwell-time distributions.

\end{abstract}

\section*{Introduction}

The human brain is intrinsically organized into large-scale networks that can be identified at rest~\cite{damois06,mantini07}. These networks  have to reorganize on a sub-second temporal scale in order to allow the precise execution of mental processes~\cite{deco11}. Spatial and temporal aspects of the dynamics underlying the reorganization of these large-scale networks require non-invasive measures with high spatial (functional magnetic resonance imaging (fMRI)) and temporal resolution (electroencephalography (EEG)). While fMRI uses the blood oxygenation level dependent (BOLD) response as a proxy for neuronal activity with a temporal resolution of several seconds, the EEG is a direct measure of neuronal activity which captures the temporal evolution of the scalp electrical field with millisecond resolution. Unlike local measures of the EEG in channel space that vary from time-point to time-point and as a function of the reference, the global measure of EEG topography remains stable for brief periods (50--100 ms) before changing to another quasi-stable state, the so-called EEG microstates~\cite{lehmann90,wackermann93}. Interestingly, four dominant topographies are consistently reported both in healthy individuals as well as in neurological and psychiatric patients at rest~\cite{koenig02, strik93, wackermann93}. While neurological and psychiatric diseases rarely affect their topography, they fundamentally alter their temporal dynamics~\cite{lehmann05, nishida13, tomescu15}. They have been coined the ``atoms of thought and can be considered the basic building blocks of spontaneous mentation that make up the spontaneous electrophysiological activity measured at the scalp~\cite{lehmann98}.

A study using simultaneously recorded EEG-fMRI identified EEG microstates as the electrophysiological correlate of four fMRI-defined resting-state networks (RSNs)\cite{britz10}. This link is surprising because EEG microstates and fMRI RSNs are two global measures of large-scale brain activity that are observed at temporal scales two orders of magnitude apart: 50--100 ms (microstates) and 10--20 seconds (fMRI-RSNs). Convolving the rapidly changing microstate time series with the hemodynamic response function (HRF) to obtain regressors for BOLD estimation acts like a strong temporal low-pass filter. The microstate time series is 'scale-free'  i.e. it shows the same behavior at different temporal scales, therefore no information is using this approach. Later, it was confirmed that this link could be established because EEG microstate time-series are mono-fractal and show long-range dependency (LRD) over six dyadic scales spanning two orders of magnitude (256 ms to 16 s)~\cite{van10}. EEG microstates and fMRI RSNs hence capture  the same underlying physiological process at different temporal scales with an electrophysiological and a hemodynamic measure, respectively. 
Importantly, Britz \emph{et al.} demonstrate that the precise timing but not the order of local transitions of the microstate sequences is crucial for their fractal properties: shuffling their local transitions without changing their timing has no effect, whereas equalizing their durations degrades the time series to white noise without memory, hence attempts at modeling microstate sequences have to go beyond modelling the local transitions. 

We note that the ``state transition process'' and the waiting (residence) time distribution for the ``renewal process'' where the transitions occur are essentially independent processes. In the case that the renewal process is memoryless then the whole process can be seen as Markov jump process, but it is possible for the transitions to be Markov but the jump times to be non-Markov.
von Wegner {\em et al.} show that neither memoryless Markov models nor single parameter LRD models fully capture the data and conclude that more sophisticated models need to be developed to understand the underlying mechanisms of microstates~\cite{von16}.

In this paper we provide a novel modelling approach based on dynamical structures called \emph{noisy network attractors}. These are stochastic models that exhibit \emph{heteroclinic} or \emph{excitable network attractors} in their noise-free dynamics \cite{ashwin16}.  A heteroclinic network is a collection of solutions (\emph{heteroclinic orbits}) that link a set of steady states (\emph{saddles}) that themselves are unstable. Excitable networks, in the sense introduced in \cite{ashwin16}, are close relations of heteroclinic networks that have a small but finite threshold of perturbation that needs to be overcome to make a transition between a number of attracting states. Heteroclinic networks have been found in models of many natural systems, for example from neuroscience~\cite{ashwin11, chossat16}, population dynamics \cite{meyer19} and game theory \cite{aguiar11}. The dynamics near a network attractor looks like a long transient: trajectories spend long periods of time close to one state before switching to another. Such transient dynamics have been observed for neural processes at a variety of levels of description \cite{Hutt17} 
and examples ranging from olfactory processing in the zebrafish~\cite{friedrich01} to human working memory~\cite{bick09} have been successfully modeled using heteroclinic cycles or networks. 
Similar networks with noise have previously been shown to produce non-Markovian dynamics~\cite{armbruster03,ashwin16}.
Given the powerful capacities of network attractors to model transient dynamics at different levels these are a promising candidate to model EEG microstate sequences.

The transition probabilities and residence times in the EEG microstates sequence are modeled using SDEs that possess such a \emph{noisy network attractor}, which is excitable (as described above) in the absence of noise. The noisy network attractor captures statistical properties of the observed microstate sequences including the distributions of residence times in each microstate and the transition probabilities. We apply this model to the analysis of EEG microstate sequences obtained from resting state EEG recordings reported in~\cite{van10}. We show that the distributions of times spent in each microstate best fit a sum of exponentials, and so the transitions are non Markov. We construct one and two layer models and apply each model to microstate sequences. We demonstrate that the one layer model produces a single exponential distribution of residence times with the same local transition probabilities as the data but no LRD. We further show that the double layer model is required to achieve distributions that are a sum of exponentials that capture both the local transition probabilities, distribution of dwell times and LRD and provides hence a better fit to the data.

\section*{Methods}
\label{sec:meth}

\subsection*{Data collection}

Detailed description of the procedures used to collect the EEG recordings and convert them into microstates are given in~\cite{van10}; here we provide a brief summary for completeness.

\subsubsection*{Subjects and Procedure}

Nine healthy volunteers (24 – 33 years, mean age 28.37 years) participated for monetary compensation after giving informed consent approved by the ethics commission of the University Hospital of Geneva. None suffered from current or prior neurological or psychiatric illness or from claustrophobia. 
For each subject, we recorded one 5-minute session outside the scanner prior to recording three 5-minute runs of resting-state EEG inside the MRI scanner. Subjects were instructed to relax and rest with their eyes closed without falling asleep and to move as little as possible. Data from one subject had to be excluded due to subsequent self-report of sleep and the presence of sleep patterns in the EEG, and the data from the remaining eight subjects were submitted to further analysis.

\subsubsection*{EEG Recording}

The EEG was recorded from 64 sintered Ag/AgCL electrodes mounted in an elastic cap and arranged in an extended 10-10 system. The electrodes were equipped with an additional 5$k\Omega$ resistor, and impedances were kept below 15$k\Omega$. The EEG was digitized using a battery-powered and MRI-compatible EEG system (BrainAmp MR plus, Brainproducts) with a sampling frequency of 5 kHz and a hardware bandpass filter of 0.016 – 250 Hz with the midline fronto-central electrode as the physical reference. The amplifier was placed ca. 15 cm outside the magnet bore and data were transmitted via fiberoptic cables to the recording computer placed outside the scanner room.

\subsubsection*{EEG Data Preprocessing}

Three types of artifacts were removed for data recorded inside the scanner: first, gradient artifacts were removed using a sliding average \cite{Allen2000} and then the EEG was down-sampled to 500 Hz and low-pass filtered with a finite-impulse response filter with a low-pass of 70 Hz. Next, the ballistocardiographic (BCG) artifact was removed using a sliding average and finally,  independent component analysis (ICA) was used to remove the residual BCG artifact along with oculomotor and myogenic artifacts. 
Data recorded outside the scanner were first downsampled to 500 Hz and subsequently, oculo-motor and myogenic artifacts were removed using ICA. Finally, both the EEG recorded inside and outside the scanner was further downsampled to 125 Hz and bandpass filtered between 1 and 40 Hz. 

\subsubsection*{EEG Microstates}

The Global Field Power (GFP) is a measure of the overall strength of the scalp electrical field. Between the local troughs of the GFP, the scalp topography remains stable and only varies in strength. Hence, the local peaks of the GFP are the best representative of an EEG microstate. We extracted the EEG at all local peaks of the GFP and submitted those to a modified Atomize-Agglomerate Hierarchical (AAHC) clustering method \cite{Tibshirani2005} in order to determine the best solution representing the most dominant topographies. The best solution was identified using a cross-validation criterion \cite{PascualMarqui1995}, a measure of residual variance, which identified four template maps as the optimal solution in each run. Finally, we computed the spatial correlation between the four dominant template maps of the optimal solution and the continuous EEG data using a temporal constraint criterion of 32 ms to obtain the time-course of the dominant EEG microstates.

In total we process and analyze 8 recordings outside the scanner and 24 recordings inside the scanner. We convert the resting state EEG into time series where each time point of the recording is classified into one of the four classically identified microstates~\cite{koenig02, strik93, wackermann93}. The raw data is first cleaned to remove oculo-motor and other artifacts, and resampled at 125Hz.
For each time point of the cleaned data we computed the global field potential 
\[
{\mathcal{G}}(t) = \sqrt{\frac{\sum^n_{i=1}(v_i(t) - \bar{v}(t))^2}{n}} 
\]
where $v_i(t)$ is the voltage of channel $i$ and $\bar{v}(t)$ is the voltage mean. The EEG topography at the local maxima of the \gfp{} has a good signal to noise ratio and so we take these points as momentary maps. All momentary maps then undergo a clustering analysis  to identify the four dominant map topographies, these are the microstates. The four microstates are shown in \fref{fig:1layer}\textbf{B}.
Finally, each time point of the (resampled) recording is classified into one of the four possible microstates.

\subsubsection*{EEG Microstate Sequence Analysis}

We denote the sequence of microstates $m(t)$ where $m\in\{1,2,3,4\}$ at any given sampling time point $t$. We classify $m(t)$ into {\em epochs} of the same state defined by saying that we enter a new epoch if and only if $t=0$ or $m(t+1)\neq m(t)$.

Analogous to the model output, we describe the $k$th epoch in terms of its {\em state} $\sigma(k)$ and {\em residence time} $\rho(k)$ respectively. This means we choose the unique sequence such that if $$\sum_{i=1}^{k-1} \rho(i)<j\leq \sum_{i=1}^{k} \rho(i)$$ then  
$
m(j)=\sigma(k)
$
and moreover $\sigma(k)\neq \sigma(k+1)$ (there is a change of state at the end of each epoch).

Specifically, we define
\begin{itemize}
	\item $R(t)$ is the distribution of residence times $\rho(k)$ for all epochs $k$.
	\item $T(m,j)$ is the probability of transition from an epoch in state $m$ to one in state $j$,
	$$
	T(m,j)= \frac{\#\{k~:~\sigma(k)=m~\mbox{ and }\sigma(k+1)=j\}}{\#\{k~:~\sigma(k)=m\}}.
	$$
\end{itemize}

We call the sequence of nodes visited $\sigma(k)$ the \emph{transition process}, and a sequence of residence times $\rho(k)$ the \emph{renewal process}.
These processes are essentially independent. The state transition process is often represented as a transition matrix containing the probabilities $T$. Each probability only depends on the state that you are currently in. In the case that the renewal process is memoryless, the whole process can be seen as Markov jump process, but it is possible for the transitions to be Markov but the renewal process to be non-Markov. In the latter case the distribution of residence times would be not be a simple exponential.

The average number of epochs per recording for outside is 2010, and for inside is 3031.
We compute $R$ by plotting a histogram of the residence times for each recording, then find the average and standard error of each bin over all recordings in that group (outside or inside). Similarly we calculate  $T$ for each recording and find the average and standard error for each probability.

\subsection*{Mathematical models}

\subsubsection*{Single-layer network model}

We aim to build a model that captures the statistical properties of the transition and renewal processes.
To this end, we construct a model of stochastic differential equations perturbed by low amplitude additive white noise, using a general method that allows us to realize any desired graph as an attracting heteroclinic or excitable network in phase space. This model has evolved from work in \cite{ashwin13} and is detailed in \cite{ashwin16} so we briefly outline the construction. 

\begin{figure}%
\centerline{\includegraphics[width=\textwidth]{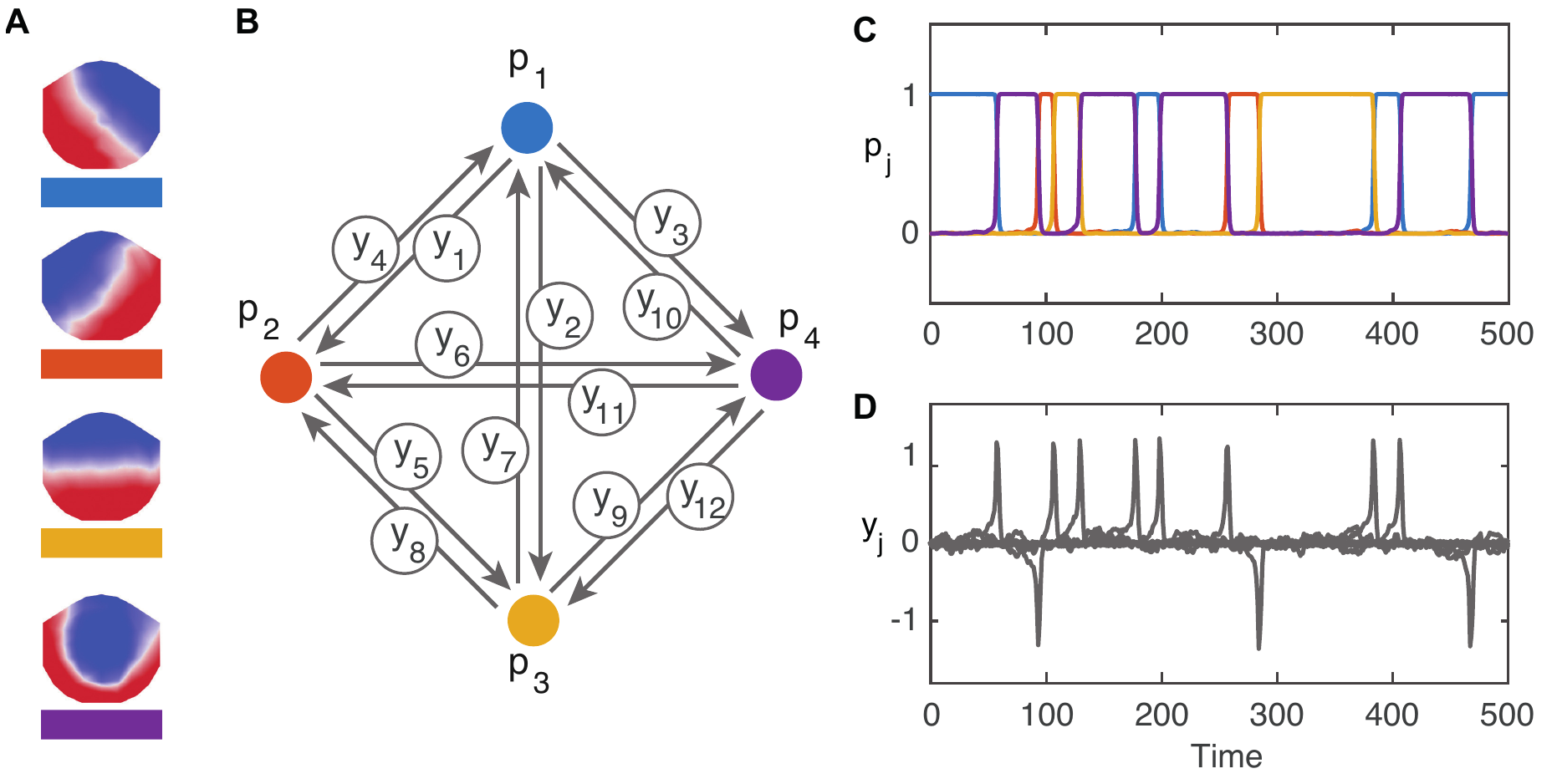}}%
\caption{\textbf{Structure and dynamics of the excitable network model.} \textbf{A} The four canonical microstates. \textbf{B} The coupling architecture of the sixteen-cell network. Each node represents one of the microstates shown in panel \textbf{A}.
	 \textbf{C} Time series of the $p$-cells (nodes), note at most only one node is equal to one at any given time point. \textbf{D} Time series of the $y$-cells (edges). The edges only become active (non-zero) during transitions between nodes.}%
\label{fig:1layer}
\end{figure}

We realize the network of all possible transitions between the four canonical microstates as an excitable network with four nodes ($p$-cells). There is an edge between two nodes in the network if there can be transition between the two corresponding microstates; here the network is all-to-all connected with twelve edges ($y$-cells). The coupling architecture of the network and corresponding microstates are shown in Figure~\ref{fig:1layer}\textbf{A--B}. 

The system is given by:
\begin{equation}
\label{eq:realiseode}
\begin{split}
\tau \dd p_{j} & = \left[ f(p_j,y)\right] + \eta_p \dd w_{j}\\
\tau \dd y_{k} & = \left[g\big(y_{k},A-B p_{\alpha(k)}^2 +C (y^2-y_{k}^2)\big)\right] \dd t + \eta_{{y}_{k}} \dd w_k
\end{split}
\end{equation}
for $j=1,\cdots,M$ and $k=1,\cdots,Q$.
The functions $f$ and $g$ are defined by
\begin{align}
f(p_j,y) &= p_j\left(F\left(1- p^2\right)+ D\left(p_j^2p^2-p^4\right)\right)+ E\left(-Z^{(o)}_j(p,y)+Z^{(i)}_j(p,y)\right), \label{eq:f}\\
g(y_{k},\lambda) &= -y_{k}\left((y_{k}^2-1)^2+\lambda\right),
\label{eq:g}
\end{align}
where 
$$p^2=\sum_{j=1}^{M} p_j^2, \ \ \ p^4=\sum_{j=1}^{M} p_j^4, \ \ \  y^2=\sum_{j=1}^{Q} y_j^2$$
and the outputs $(O)$ and  inputs $(I)$ to the $p$ cells from the $y$ cells are:
\begin{equation}
\begin{split}
Z^{(O)}_j(p,y) &= \sum_{\{k~:~\alpha(k)=j\}} y_k^2p_{\omega(k)}p_j\\
Z^{(I)}_j(p,y) &= \sum_{\{k'~:~\omega(k')=j\}} y_{k'}^2p_{\alpha(k')}^2.
\end{split}
\end{equation}
 The $w$ are independent identically distributed noise processes,  $\eta$ are noise weights and $A,B,C,D,E,F$ are constants.
Here we introduce the time scaling $\tau$ because although the $p$ cells can be scaled by the parameters the $y$ cells have a functional form which is fixed.
The full equations are given in \nameref{S1_Appendix}.

The $p_j$ variables classify which node of the network (i.e. which of the four microstates) is visited. In the system with no noise, there are equilibrium solutions with $p_j=1$ and all other coordinates zero. The $y_k$ variables become non-zero during a transition between nodes.

We use the parameter set
\begin{equation}
A=0.5,~B=1.49,~C=2,~D=10,~E=4,~F=2
\label{eq:systemparams}
\end{equation}
throughout as in \cite{ashwin16}, where $B<1.5$ gives an excitable network.
In an excitable network nodes are stable equilibria and transitions between nodes are driven by the additive noise.
The transition rates between nodes are modulated by the noise levels on the edges (rather than on the nodes) as described in \cite{ashwin16}.
We choose the noise levels so that the model statistics, namely the transition probabilities the distribution of the renewal process (residence times) fit the data; see  Model simulation and fitting for details. 
We fix the noise on the nodes $\eta_p=10^{-4}$.
The time scaling constant $\tau$ is set to 8ms (in line with the sampling rate of the data) throughout.

Figures~\ref{fig:1layer}\textbf{C}--\textbf{D} show example time series output for the nodes and the edges, respectively, for the excitable network model. At any given moment in the time series the trajectory in the simulation is close to one of the equilibrium solutions, and one of the $p_j$ variables is close to 1. To determine where the trajectory is close to a given node we define a box in phase space around that node so that when the trajectory is in the box we say it is near the node. Here we fix the box size $H= 0.49$ such that each box contains one node and the boxes do not overlap. 
The duration of time spent in the box is then the residence time for that node \cite{ashwin16}.

The output of the model simulation is a series of $k$ epochs where each epoch is defined by the state it is in $\sigma\in\{1,2,3,4\}$ and its residence time $\rho$.
The transition and renewal processes generated by the network construction given here are both Markov. 
Due to the evidence of multi-scale behavior of the temporal dynamics of microstate sequences~\cite{gschwind15, van10} we also present a more sophisticated model a with a multi-layer construction that will generate a non-Markov renewal process.

\subsubsection*{Multi-layer network model}

We construct a system of $N$ levels, where each level $l$ has $M_l$ nodes, and we assume all-to-all connections, so we thus have $Q_l\equiv M_l(M_l-1)$ edges. In each level, we set up a system of SDEs in the form of that described in~\cite{ashwin16}, as follows, where $p_{l,j}\in\R^{M_l}$ and $y_{l,k}\in\R^{Q_l}$:
\begin{equation}
\label{eq:multisde}
\begin{split}
\tau \dd p_{l,j} & = \left[f_{l}(p_j,y)\right] \dd t+\eta_p \dd w\\
\tau \dd y_{l,k} & = \left[g_{l}(y_{k},p_{\alpha(k)})+ z_{l,k}\right] \dd t+\eta_{{l,k}}  \dd w_k.
\end{split}
\end{equation}
Here we allow for a general input into the $y$-cells $z_{l,k}(t)$ that linearly couples layers from the $p_l$ nodes to the $y_{l+1,k}$ edges by:
\[
z_{l+1,k}= \sum_{j=1}^{M_{l}}\zeta_{l,j} \, p_{l,j}^2.
\]
The parameter $\zeta_{l,j}$ is a constant that scales the activity from node $p_{l,j}$ to all the edges in level $l+1$. The functions $f$, $g$, $Z^{(O)}$ and $Z^{(I)}$ are analogous to those in the single layer model,

\begin{align*}
f_{l}(p_j,y) & =  p_{l,j}\left(F(1- p_l^2)+ D(p_{l,j}^2p_l^2-p_l^4)\right) + E\left(-Z^{(O)}_{l,j}(p,y)+Z^{(I)}_{l,j}(p,y)\right), \\
 g_{l}(y_k,p_{\alpha(k)}) & = -y_{l,k}\left((y_{l,k}^2-1)^2+A - B p_{l,\alpha(k)}^2 +C(y_l^2-y_{l,k}^2)\right),
\end{align*}
for $j=1,\dots,M_l$ and $k=1,\dots,Q_l$,
\begin{equation}
\begin{split}
Z^{(O)}_{l,j}(p,y) &= \sum_{\{k~:~\alpha(k)=j\}} y_{l,k}^2p_{l,\omega(k)}p_{l,j}\\
Z^{(I)}_{l,j}(p,y) &= \sum_{\{k'~:~\omega(k')=j\}} y_{l,k'}^2p_{l,\alpha(k')}^2,
\end{split}
\end{equation}
and 
\[
p_l^2=\sum_{j=1}^{M_l} p_{l,j}^2,  \ \ \ p_l^4=\sum_{j=1}^{M_l} p_{l,j}^4, \ \ \ y_l^2=\sum_{k=1}^{Q_l} y_{l,k}^2.
\]

We consider in particular a two layer model where $M_1=2$ and $M_2=4$ and the constants $A,B,C,D,E,F$ are set as \eqref{eq:systemparams} for each layer. Note that both layers are excitable networks. Layer one with two nodes is a ``hidden layer" and we only consider the output from layer 2 with four nodes. As before the output is a transition process and a renewal process.

\begin{figure}[th]%
\centerline{\includegraphics[width=\textwidth]{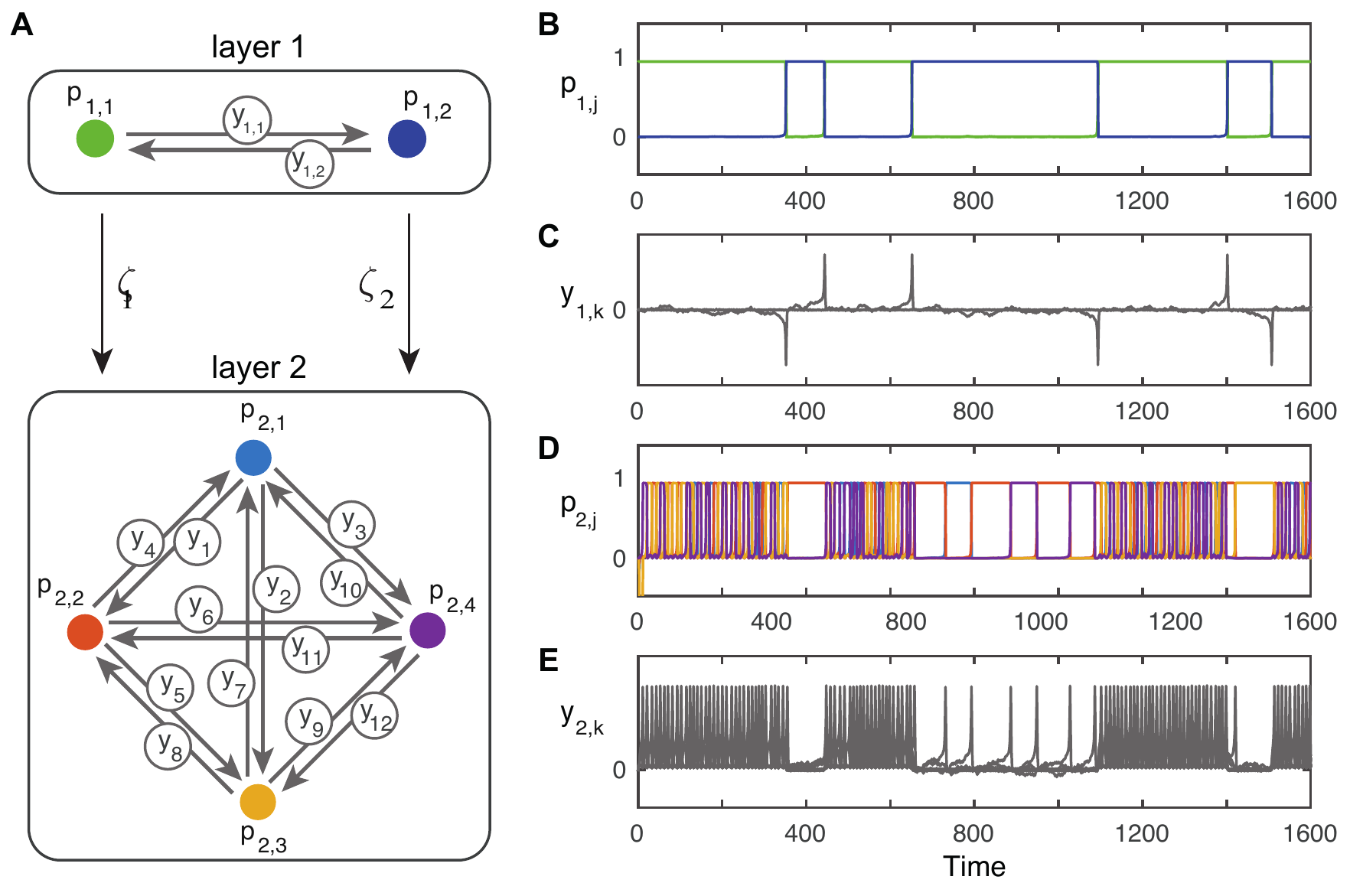}}%
\caption{%
	\textbf{Structure and dynamics of a two-layer model.} \textbf{A} The coupling architecture of the two networks. Compare layer two to \fref{fig:1layer}\textbf{B}. Example time series are shown for layer one nodes \textbf{B} and edges \textbf{C}, layer two nodes \textbf{D} and edges \textbf{E}. The noise on all edges $\eta_{l,k}=10^{-2}$ is fixed, residence times are scaled  by $\zeta_1 = 0.1$ and $\zeta_2 = 0.001$. When $p_{1,2}=1$ in panel \textbf{A} the noise on the edges in layer two is scaled by $\zeta_2$ and residence times of nodes in layer two, shown in panel \textbf{D}, are much longer than when $p_{1,1}=1$. }%
\label{fig:2layer}%
\end{figure}

\Fref{fig:2layer}\textbf{A} shows the coupling architecture for the two layer model, \textbf{B--C} show the time series of the nodes and edges in layer one, respectively, and \textbf{D--E} show the time series of the nodes and edges in layer two, respectively.
Compare layer two topography and dynamics to \fref{fig:1layer}.
The two nodes in layer one effect the dynamics on the edges (and therefore the residence times) in layer two by the scalings $\zeta_{1,1}=\zeta_1$ and $\zeta_{1,2}=\zeta_2$. These scale the dwell times in the renewal process output from layer two. 
For illustrative purposes we choose $\zeta_{1}=10^{-1}$ and $\zeta_{2}=10^{-3}$ here. Panels \textbf{B--E} clearly show that when $p_{1,2}=1$ the residence times at each node $p_{2,j}$ are longer (transitions between nodes are less frequent) as the edges in layer two $y_{2,k}$ are scaled by $\zeta_{2}$.
Note that if $\zeta_{1}=\zeta_2$ layer one would be redundant as the residence times would be consistent (drawn from the same distribution) in either node and the renewal process would again be Markov.

\subsection*{Model simulation and fitting \label{sec:fit}}

We wish to use the model capture the transition probabilities and distribution of residence times of the data.
To this end we first fit exponential curves to the residence distributions for the data recorded inside and outside the fMRI scanner.
Specifically, we compute the histogram of residence times from the microstate sequences for bin size $40$. The distribution is truncated at the first empty bin. We then use the {\tt{fit}} function in {\sc{Matlab}} to fit the single and double exponentials
\begin{equation}
\label{eq:decayfit}
{\mathcal{E}}^{(1)}(t)= A_1\exp(-k_1t),~~{\mathcal{E}}^{(2)}(t)= A_1\exp(-k_1t)+A_2\exp(-k_2t),
\end{equation}
to each dataset where we constrain $A_i>0$ and $k_i\geq 0$.
We use an F-test to indicate whether the double exponential was a better fit to the data; code used as written for~\cite{ftest}. 

We numerically simulate the one and two-layer models with a stochastic Heun method implemented using a custom code written in {\sc{Matlab}}. 
We compute 10 realizations of the model using step size $0.05$ up to maximum of 100,000 steps. This gives approximately 5750 epochs per realization. We calculate the residence time distribution $R$ and transition probabilities $T$, then average the values over 10 realizations.
The results of the averaged simulations (model output) are compared by eye to the data. 

To fit the one-layer model output to the data we adjust the  noise amplitude parameters on the edges $\eta_{y_k}$.
The noise amplitude on each edge controls the probability of transition along that edge, therefore we change the amplitudes so the model output transition probabilities are within the error bars of the transition probabilities from the data. For example if $T(1,2) > T(2,1)$ then we set $\eta_{y_1}>\eta_{y_4}$.
The overall magnitude of the $\eta_{y_k}$ parameters controls the decay rate of the residence time distribution. If all $\eta_{y_k}$ are large, for example, $O(10^{-1})$ transitions happen quickly, the residence times are short, and the decay rate of the distribution $R$ is large (slope is steep); whereas if $\eta_{y_k}$ are small, for example, $O(10^{-4})$ there are long residence times between transitions and the decay rate is small (slope is shallow).
In this way we can fit the distribution of the model to the fitted curves ${\mathcal{E}}^{(1)}(t)$.

To fit the transition probabilities in the two-layer model we adjust the noise values of the edges in layer two $\eta_{2,k}$ in the same way as for the one-layer model. To fit the residence distributions we first fix the overall magnitude of the $\eta_{2,k}$ in line with the one-layer model, then adjust the transfer parameters $\zeta_j$ and noise values of the edges on layer one $\eta_{1,k}$.
Changing $\zeta_j$ changes the dynamics on the edges of layer two in a homogeneous way (the same scaling on all edges): if $\zeta$ is increased the residence times decrease and the decay rate of the distribution is becomes larger; if $\zeta$ is decreased, the decay rate of the distribution of the residence times is decreased. As there are two transfer parameters (one for each node in layer one) the residence distribution of the output from the model is a linear combination of the two decay rates; one $\zeta$ is associated with the distribution at short times and the other captures the heavy tails.
The proportion of each distribution (the mixing) is controlled by the noise on the edges is layer one.
If $\eta_{1,1}>\eta_{1,2}$  more time will be spent in node $p_{1,2}$ than in node $p_{1,1}$ so in the residence distribution there will be a larger proportion of the decay rate associated with $\zeta_2$.
In this way we can fit the distribution of the model to the fitted curves ${\mathcal{E}}^{(2)}(t)$.

\section*{Results}

\subsection*{Microstate residence time distributions have two phase decay}

\begin{figure}[t]
\center {\includegraphics[width=\textwidth]{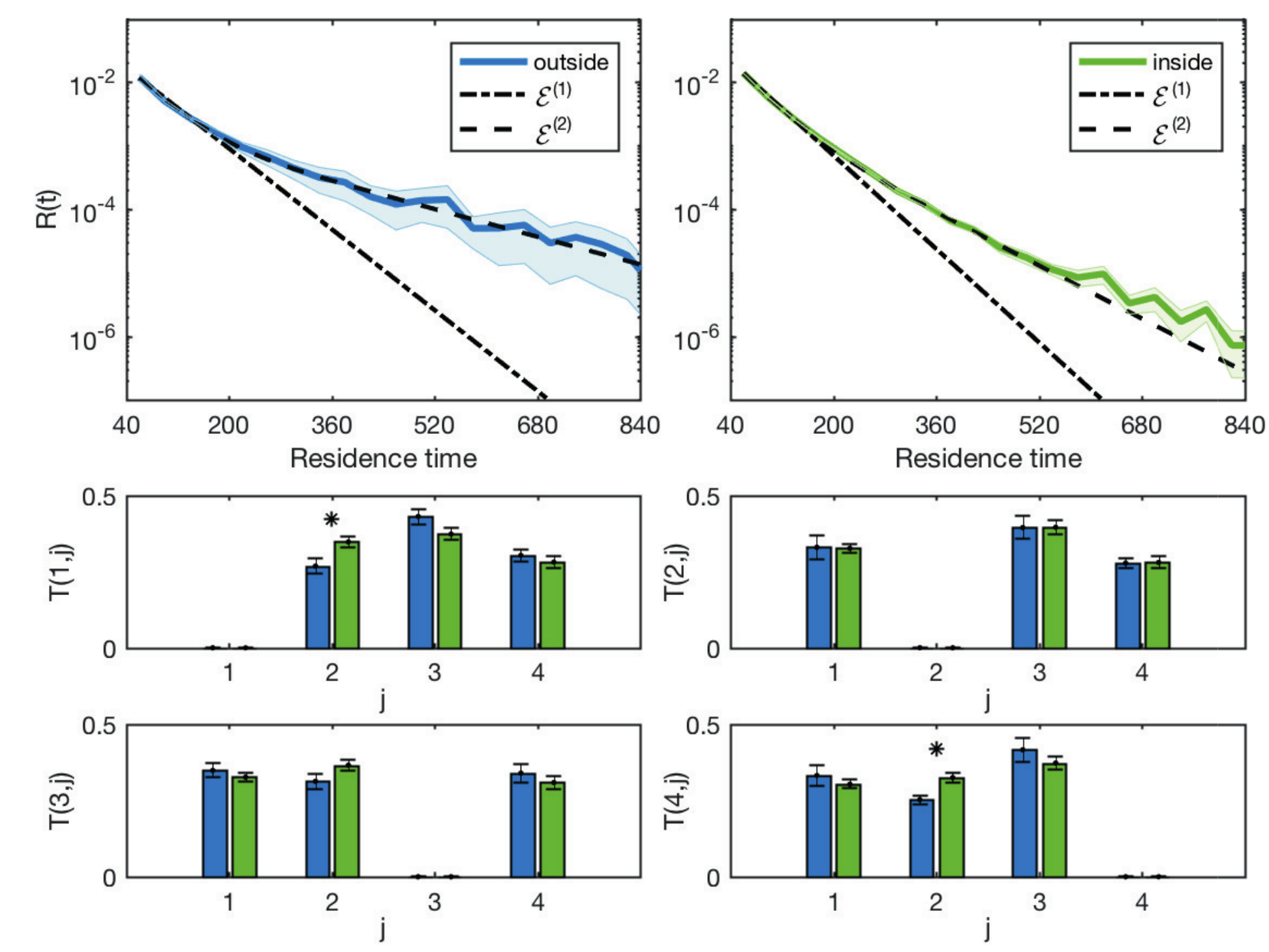}}
\caption{\textbf{Statistical properties of the recordings and curve fitting.} 
Top row shows histograms of residence times for the microstate sequences outside and inside the scanner, with bin size 40. Note the data is truncated at the first empty bin. The best fit curves for one-phase decay ${\mathcal{E}}^{(1)}$ and two-phase decay ${\mathcal{E}}^{(2)}$ are shown for each distribution. 
The bottom four panels show the probability of transition $T(m,j)$ from state $m$ to state $j$  for each microstate $m={1,2,3,4}$. Significant differences $p<0.05$ are denoted with a star. }
\label{fig:msstats}
\end{figure}

The residence time distributions $R(t)$ for data collected inside and outside the scanner and the transition probabilities $T(m,j)$ for the EEG recordings are shown in \Fref{fig:msstats}.
The distributions $R(t)$ are plotted on a logarithmic scale with ${\mathcal{E}}^{(1)}(t)$ and ${\mathcal{E}}^{(2)}(t)$ given by \eqref{eq:decayfit}.
The transition probabilities for each dataset are shown in panels \textbf{C}.

For each dataset we perform a comparison of fits of ${\mathcal{E}}^{(1)}$ and ${\mathcal{E}}^{(2)}$ to the data. We find that an $F$-test rejects the null hypothesis that the distribution is ${\mathcal{E}}^{(1)}$: details are given in Table~\ref{tab:residenceFtest}.
Using the Kruskall-Wallis ANOVA the mean rank fo the residence time distributions are significantly different ($p<0.0005$).
Significant differences in transition probabilities between outside and inside can be seen between in the transition from state 1 to state 2, and from state 4 to state 2. 
Note there are no self transitions due to our definition of an epoch.

\begin{table}[h]
\centering
\caption{
{\bf Two phase decay captures residence time distributions.}}
$$
\begin{array}{r|rr}
&\multicolumn{2}{c}{{\mathcal{E}}^{(1)}}\\
&	{\rm Outside}	&	{\rm Inside}	\\
\hline									
A1	&	3.523 \times 10^{-2}	&	4.931 \times 10^{-2} \\
k1	&	1.828 \times 10^{-2}	&	2.119 \times 10^{-2} \\
&\multicolumn{2}{c}{{\mathcal{E}}^{(2)}}\\
&	{\rm Outside}	&	{\rm Inside}	\\
\hline	
A1	&   4.111 \times 10^{-2}	&	5.141 \times 10^{-2}	\\
k1	&	2.340 \times 10^{-2}	& 	2.620 \times 10^{-2}	\\
A2	&	2.617 \times 10^{-3} 	&	6.611 \times 10^{-3} 	\\
k2	&	6.249 \times 10^{-3}	&	1.198 \times 10^{-2}	\\							
\hline									
F(d_n,d_d) & 5685(2,32) 	& 1105(2,32)	 \\
\mbox{Preferred} & \mathcal{E}^{(2)} & \mathcal{E}^{(2)}
\end{array}
$$
\begin{flushleft} 
Best fit parameters for one-phase decay ${\mathcal{E}}^{(1)}$ and two-phase decay ${\mathcal{E}}^{(2)}$  as in (\ref{eq:decayfit}).  The final rows give the F-test results with threshold $\alpha=0.05$.
\end{flushleft}
\label{tab:residenceFtest}
\end{table}

\subsection*{One-layer network model captures transition probabilities}

\begin{figure}[th]
	\centering
	\includegraphics[width=\textwidth]{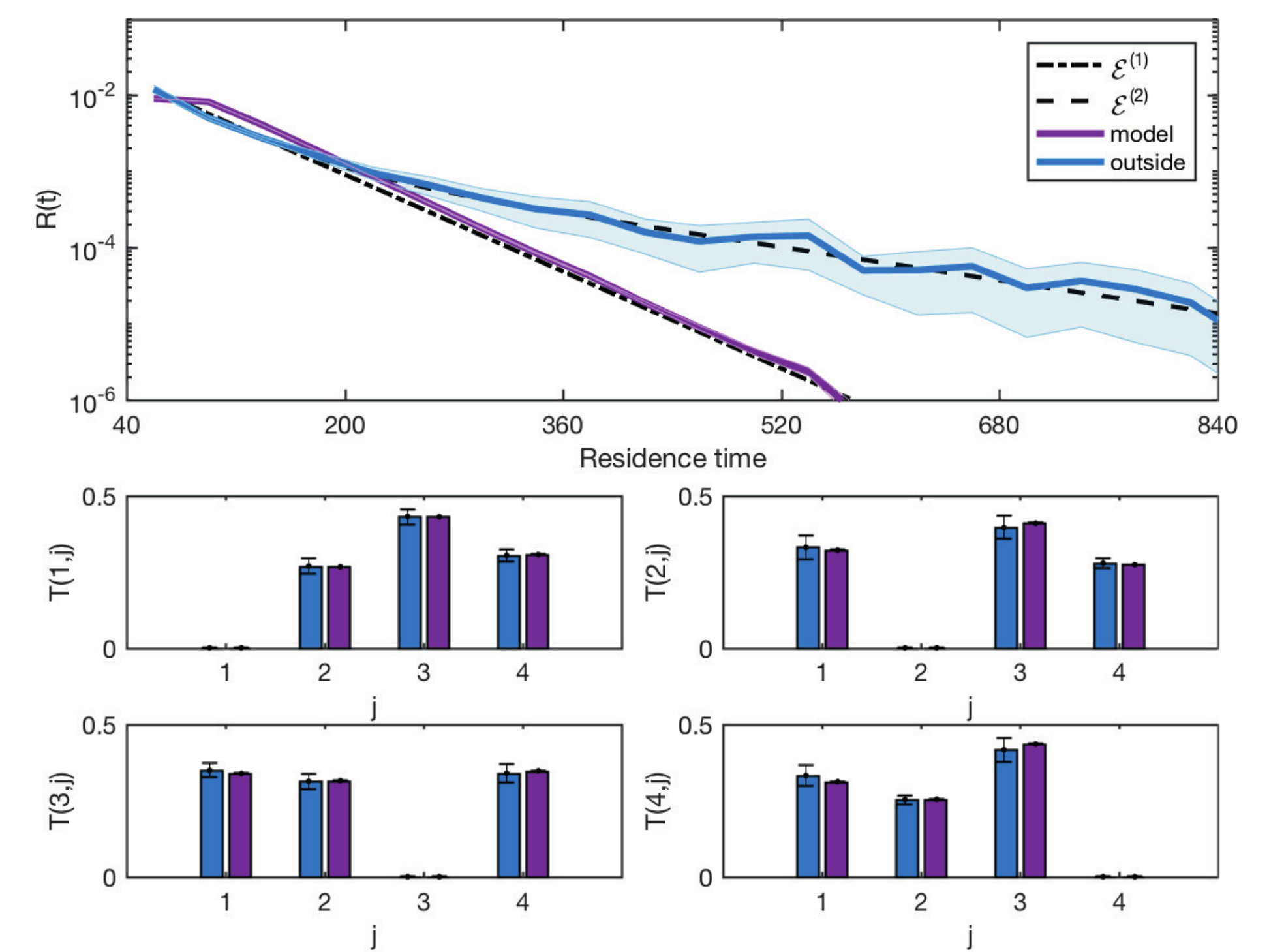}
	\caption{\textbf{One-layer model captures transition probabilities.} One-layer model simulation (purple) with residence distribution for the `outside' dataset (blue) in the top panel. The single and double exponential fits from~\tref{tab:residenceFtest} are shown for comparison (black dashed lines). The transition probabilities $T(m,j)$ are shown below for both the data (blue) and the simulation (purple). The parameters used in the simulations in each panel are given in~\tref{tab:1lfit}. There is good agreement between the model transition probabilities and the data as this was our fitting criteria. The residence time distribution of the model closely follows the single exponential fit to the data but does not capture the data distribution well at all. }
	\label{fig:1lfit}
\end{figure}

The results of fitting the one-layer model to the data from outside the scanner are shown in \fref{fig:1lfit}. The single exponential fit ${\mathcal{E}}^{(1)}$ from~\tref{tab:residenceFtest} is shown. The transition probabilities for the model and the data show a good fit as the model probabilities are within the error bars of the data for all transitions.  The noise weights $\eta_{y_k}$ used are given in \tref{tab:1lfit}.
The distribution of the one-layer model aligns with ${\mathcal{E}}^{(1)}$. However,  this model fails to capture the two decay rates observed in the data.

\begin{table}[h]
	\caption{\textbf{Parameters for the one-layer model.}}
	$$
	\begin{array}{rcccccc}
		\hline
		 \eta_{y_1}&    \eta_{y_2}&    \eta_{y_3}&  \eta_{y_4}&   \eta_{y_5}&  \eta_{y_6}\\
		0.0272 &  0.0324 & 0.0286 &  0.0301 &  0.0329 &  0.0284 \\
		 \eta_{y_7}&    \eta_{y_8}&    \eta_{y_9}&  \eta_{y_{10}}&   \eta_{y_{11}}&  \eta_{y_{12}}\\
		0.0276 & 0.0269 &0.0278 &   0.0290 &   0.0270 &  0.0328  \\
		\hline
	\end{array}
	$$
	\begin{flushleft} 
	The noise values on the edges $\eta_{y_{k}}$ are given for the model simulation to fit the `outside' dataset shown in \fref{fig:1lfit}.
	\end{flushleft}
	\label{tab:1lfit}
\end{table}

\subsection*{Two-layer model captures two-phase decay of residence times}

\begin{figure}[th]
	\centering
	\includegraphics[width=\textwidth]{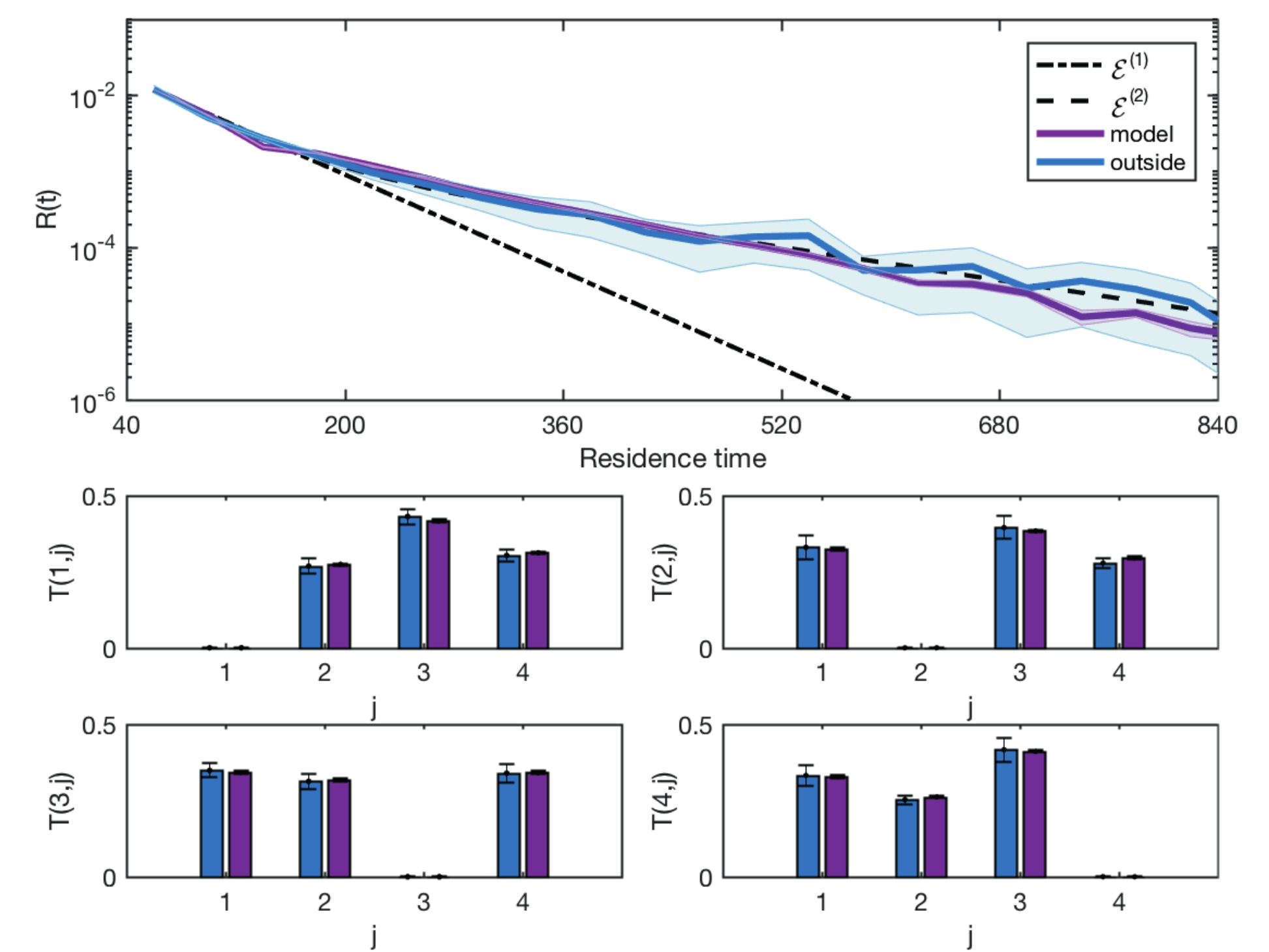}
	\caption{\textbf{Two-layer model captures longer residence times.} The distribution for the `outside' dataset (blue) with the residence time distributions from the simulation of the two-layer model (purple); see \fref{fig:2layer}. The parameters used in the simulations in each panel are given in~\tref{tab:2lfit}. The double exponential fit ${\mathcal{E}}^{(2)}$  from~\tref{tab:residenceFtest} is also shown. The two layer model fits the transition probabilities well as this was the fitting criteria. The model simulations fit the two decay rates of the residence time distribution of the data capturing both long and short times.}
	\label{fig:2lfit}
\end{figure}

The results of fitting the two-layer model to the data from outside the scanner are shown in \fref{fig:2lfit} with the double exponential fit ${\mathcal{E}}^{(2)}$ from~\tref{tab:residenceFtest}. The distributions of the simulations agree closely with ${\mathcal{E}}^{(2)}$ and fit within the error bars of the distribution. The transition probabilities for the model and data are also shown. 
The parameters for the simulations are shown in~\tref{tab:2lfit}.
The two layer model captures the two-phase decay of the data.

\begin{table}[h]
	\caption{\textbf{Parameters for the two-layer model.}}
	$$
	\begin{array}{cccccc}
	\hline
	\zeta_{1}&    \zeta_{2}& \eta_{1,1} &  \eta_{1,2}&  &  \\
	0.19 &  0.0001&   0.05 &  0.05 && \\
	\hline
	 \eta_{2,1}&    \eta_{2,2}&    \eta_{2,3}&  \eta_{2,4}&   \eta_{2,5}&  \eta_{2,6}\\
	   0.00080 &  0.00230 & 0.00120 &  0.00150 &  0.00210 &  0.00110 \\
	  \eta_{2,7}&    \eta_{2,8}&    \eta_{2,9}&  \eta_{2,10}&   \eta_{2,11}&  \eta_{2,12}\\
 	 0.00138 & 0.00120 &0.00138 &   0.00146 &   0.00080 &  0.00250  \\
	\hline
	\end{array}
	$$
	\begin{flushleft} 
		The noise values on the edges $\eta_{l,k}$ are given for each layer $l=1,2$ with the transfer parameters $\zeta_n$ for the model simulation shown in \fref{fig:2lfit}. Compare the noise values on the edges of layer two $\eta_{l,k}$ to the values in \tref{tab:1lfit}. 
	\end{flushleft}
	 \label{tab:2lfit}
\end{table}

\section*{Discussion}

In this article we have demonstrated a new modelling approach applied to capture the statistical properties of the temporal dynamics of EEG microstates.
We analyze EEG microstate sequences from EEG data recorded inside and outside an fMRI scanner, previously reported in~\cite{britz10,van10}.
We consider the transition probabilities between microstates and the distributions of the residence times.

We show that there are significant differences between the residence time distributions.
Both residence time distributions are best fit by two decay rates. We note the similarity in the first decay rates given by $2.340 \times 10^{-2}$ and $2.620 \times 10^{-2}$ for outside and inside respectively. However, the second decay rate is very different $6.249 \times 10^{-3}$ and $1.198 \times 10^{-2}$ for outside and inside respectively.
We also see a change in the transition probabilities; the inside dataset transition probabilities have less variance than the outside dataset, with a significant increase of the probability of transition to microstate 2 from microstates 1 and 4. 
The most likely explanation for the difference in long-range temporal correlation for the data recorded inside and outside the scanner is the difference in artifact removal.
The data recorded outside the scanner were only contaminated by very occasional oculomotor artifacts (eye-movements) which can be reliably detected and removed with ICA. EEG data recorded inside the scanner are additionally contaminated by the gradient and the ballistocardiographic artifacts. The gradient artifact has an amplitude three orders of magnitude larger than the EEG, but it is very regular and can thus be removed reliably by a sliding average. The amplitude of the ballistocardiogram is roughly an order of magnitude larger than the EEG, and its removal requires two steps: a sliding average to remove the major amplitude and ICA to remove the residual. The latter is essential but also prone to remove not only the residual artifacts but also to remove part of the signal. 

In the past few years there have been various attempts to model EEG-microstate sequences. G{\"a}rtner {\em et al.} \cite{gartner15} construct a hidden-Markov type stochastic model based on the transition probabilities between microstates extracted from the data. The transition matrix gives the probabilities of moving from one state to another, where each probability only depends on the state that you are currently in. Such a one-step Markov construction assumes that the transitions between microstates depend on the current state but otherwise are memoryless. However, no restrictions are placed on the distributions of the residence times. 

The Markov model approach taken in \cite{gartner15} has been criticized for the underlying assumption that the microstate transition process is independent of the underlying global field power time series~\cite{koenig15}. Further, the authors do not comment on possible long range dependencies (LRDs) in the data. Gschwind {\em et al.}  argue that LRDs are an intrinsic property of neural dynamics that manifest in the microstate sequences as scale-free or fractal structure in the data~\cite{gschwind15}. To elucidate this fractal structure and verify and expand the work of \cite{van10} they subject the data to a battery of tests, including computing the power spectral density and Hurst exponents using detrended fluctuation analysis (DFA), a wavelet framework, and time-variance analysis.
Using wavelet detrended fluctuation analysis EEG-microstate time series have been shown to exhibit mono-fractal behavior over six dyadic scales indicative of long range  ~\cite{britz10}.

Further analysis of scale-free properties in EEG-microstate data, subsequently conducted by von Wegner {\em et al.}, shows that these measures, when applied to real (instead of simulated) EEG-data, do not guarantee the existence of LRDs~\cite{von16}. Specifically, they compute the Hurst exponent of a given EEG-microstates data set using three different methods and perform statistical tests that show that the results are not always significantly different from a constructed white noise process. The assumptions underlying the analysis that provide a connection between the Hurst exponent and long range dependencies in data and therefore temporal fractal structure, may also not be applicable to real EEG-data.

That the microstate residence times presented here are best represented by two decays rates provides additional evidence that the temporal properties of microstates are non-Markovian in line with \cite{gschwind15}. 
It is clear that Markovian or single parameter LRD models do not adequately represent the temporal dynamics of microstate sequences \cite{von16}.

One limitation of this approach is that we only consider memoryless transitions, and assume that the transition probabilities do not change over time. These assumption may be incorrect as they do not fit with the non-Markovian properties of microstate sequences previously found~\cite{van10}. Future work would be to look at transition probabilities of sequences of microstates, thereby including different levels of history.

Combined fMRI-EEG studies have found correlates between each microstate and resting state networks~\cite{van10}; A is the auditory network, B is the visual network, C is the saliency network and D is the attention network.
The switching dynamics between microstates indicates changes in switching between these underlying neural networks~\cite{britz10}. 
Multiple decay rates point to complex switching dynamics required for cognition. 
More work needs to be done to understand how these dynamic changes correlate to the cognitive dysfunction observed in this state, for example loss of short term memory.

Finally this modelling approach could be used to identify dynamic differences and their brain network correlates at different levels of consciousness and in other neurological disorders for example Alzheimer's disease, epilepsy and schizophrenia.


\section*{Conclusions}

Using EEG data from~\cite{van10} we show that the residence time distributions of resting state EEG microstate sequences have a two phase decay.
We also show that the additional processing of the data recorded inside the fMRI scanner leads to a decrease of very long residence times ($>900$ms) in the distribution.
Using the construction outlined in \cite{ashwin16} we build systems of stochastic differential equations (SDE) that have a ``noisy network attractor''. Using a one-layer model we show that this captures the transition probabilities between microstates. The distribution given by this model is (single) exponential.
We also build a two-layer model with a (hidden) layer containing two nodes each associated with a transfer parameter that scales the behavior of the edges in the four-node network layer. This model produces a residence time distribution with two decay rates (sum of two exponentials) that better captures the two-phase decay seen in the data.
The identification of EEG-microstates as correlates of resting state brain networks means that analysis of the temporal dynamics of microstate sequences can lead to crucial insights into the switching between these networks required for cognition and perception in resting state and could provide indicators of underlying mechanisms of dynamic changes characteristic of neurological disorders.

\section*{Supporting information}

\paragraph*{S1 Appendix.}
\label{S1_Appendix}
{\bf Single-layer model equations.}

\section*{Acknowledgements}

JC and PA gratefully acknowledge the financial support of the EPSRC Centre for Predictive Modelling in Healthcare, via grant EP/N014391/1. 
JC acknowledges funding from MRC Skills Development Fellowship MR/S019499/1.
CMP acknowledges funding from the Marsden Fund, Royal Society of New Zealand.


\end{document}